# Orientation selection of equiaxed dendritic growth by three-dimensional cellular automaton model

Lei Wei, Xin Lin, Meng Wang, Weidong Huang

1 State Key Laboratory of Solidification Processing, Northwestern Polytechnical University, Xi'an 710072 P.R. China

**Abstract**

A three-dimensional (3-D) adaptive mesh refinement (AMR) cellular automata (CA) model is developed to simulate the equiaxed dendritic growth of pure substance. In order to reduce the mesh induced anisotropy by CA capture rules, a limited neighbor solid fraction (LNSF) method is presented. An expansion description using two interface free energy anisotropy parameters ($\varepsilon_1$, $\varepsilon_2$) is used in present 3-D CA model. The dendrite growths with the orientation selection between <100> and <110> are discussed using the different $\varepsilon_1$ with $\varepsilon_2$=-0.02. It is found that the simulated morphologies by present CA model are as expected from the minimum stiffness criterion.

**Keywords: cellular automaton; solidification; dendrite; anisotropy**

**Introduction**

The solidification microstructures, which generally determine the final properties of the relative product to a large extent [1, 2], have been focused by the physicists and material scientists for several decades. Dendrite is the most common microstructures observed in the casting products of commercial alloys [3-8].

According to the solvability theory [3-6], a dendrite tip is determined by a stability parameter, which is a function of the anisotropy parameter of the interface free energy. Recently, molecular dynamics (MD) simulations [9] have established that the parameterization of interface free energy for fcc metals requires two anisotropy parameters ($\varepsilon_1$, $\varepsilon_2$).

Phase-field simulations [10] have showed that dendritic preferred growth directions vary continuously from <100> to <110>, when the anisotropy parameter $\varepsilon_1$ is reduced from 0.15 to 0.0 while the anisotropy parameter $\varepsilon_2$ is fixed to -0.02. However, for a large region sandwiched between the <100> and <110> regions (0.05<$\varepsilon_1$<0.12, $\varepsilon_2$=-0.02), the phase-field results are not as expected from the minimum stiffness criterion.

Cellular Automaton (CA) model [11-21] is another computational model for describing dendritic growth, and it has been used to simulate many solidification phenomena, such as columnar-to-equiaxed transition (CET) [15] and the dendritic growth under forced flow [16].

Pan [17] presented a 3-D cellular automaton model for describing solutal dendrite growth, in which a Cahn-Hoffman vector is used to calculate the capillarity undercooling. In his model, only dendrite growth with <100> preferred direction could be simulated, because only one anisotropy parameter $\varepsilon_4$ is used. However, he did not quantitatively examine the effects of $\varepsilon_4$ on the dendrite growth behaviors.

Due to the mesh induced anisotropy, few of the CA models [11-18] have quantitatively investigated the effects of interface free energy anisotropy on dendritic growth. It is well known that the cellular automaton rule could lead to a mesh induced anisotropy in dendrite growth, even when the interface free energy is isotropic [19-21].

Based on the previously developed 2-D CA model [20, 21], which could lower the mesh induced anisotropy to a large extent, a 3-D CA model with lower mesh induced anisotropy is presented in present work. Since it is difficult for the 3-D CA model to use the 2-D zigzag capture rule developed in the previous model [20, 21], a limited neighbor solid fraction (LNSF) method is developed for the 3-D CA model to reduce the mesh induced anisotropy.

## 1. Model Description

### 1.1 Thermal field

During 3-D equiaxed dendritic growth, the thermal field is governed by the following equation:

$$\frac{\partial T}{\partial t} = \alpha \left( \frac{\partial^2 T}{\partial x^2} + \frac{\partial^2 T}{\partial y^2} + \frac{\partial^2 T}{\partial z^2} \right) + \frac{L}{C_p} \frac{\partial f_s}{\partial t} \tag{1}$$

where $T$ is temperature, $t$ is time, $\alpha$ is the heat diffusivity, $L$ is latent heat of solidification, $C_p$ is specific heat and $f_s$ is solid fraction. To solve Equation (1), a central finite difference method was used.

### 1.2 Anisotropic interface free energy and interface stiffness

The atomistic calculations [9] have established that the parameterizations of anisotropic interface free energy require two anisotropy parameters, as expressed by Eq. (2)

$$\frac{\gamma(\hat{n})}{\gamma_0} = 1 + \varepsilon_1 \left( \sum_{i=1}^{3} n_i^4 - \frac{3}{5} \right) + \varepsilon_2 \left( 3 \sum_{i=1}^{3} n_i^4 + 66 n_1^2 n_2^2 n_3^2 - \frac{17}{7} \right) + \ldots \tag{2}$$

where $\gamma_0$ is the averaged interface free energy, $\hat{n} = (n_1, n_2, n_3)$ is the interface normal, $\varepsilon_1, \varepsilon_2$ are the two anisotropy parameters.

According to extremum stiffness criterion, the minima of the interface stiffness, $S = \gamma + d^2\gamma/d\theta^2$, which appears in the Gibbs-Thomson condition, indicates the dendrite growth directions [10]. The interface stiffness expressed in spherical angular coordinates is [10].

$$S = 2\gamma + \frac{\partial^2 \gamma}{\partial \theta^2} + \frac{1}{\sin^2 \theta} \frac{\partial^2 \gamma}{\partial \varphi^2} + \cot \theta \frac{\partial \gamma}{\partial \theta} \tag{3}$$

where, $\theta$ and $\varphi$ are the spherical angles of the interface normal. To solve Eq. (3), it is needed to substitute $\hat{n} = (\cos\theta\sin\varphi, \sin\theta\sin\varphi, \cos\varphi)$ into Eq. (2). Then, the plot of $1/S$ could be draw for different anisotropy parameters.

### 1.3 Gibbs-Thomson equation and the weighted mean curvature

Hoffman and Cahn [22] introduced a $\xi$-vector to represent the anisotropic interface free energy of a sharp interface. The definition of Hoffman-Cahn $\xi$-vector is as follow.

$$\xi \equiv grad(r\gamma) \tag{4}$$

where, $r$ is the distance from the interface to the coordinate origin, $\gamma$ is the interface free energy as a function of interface normal $\hat{n}$.

According to Taylor's work [23], the capillarity undercooling can be related to the weighted mean curvature (*wmc*), which is equivalent to the negative of the surface divergence of the $\xi$-vector.

$$wmc = \sum_{i=1,2}\left[\gamma(\hat{n}) + \frac{\partial^2 \gamma(\hat{n})}{\partial \theta_i^2}\right]\frac{1}{R_i} = -div_S \xi \tag{5}$$

Thus, the Gibbs-Thomson equation based on the weighted mean curvature can be expressed as.

$$T_I = T_M - \frac{T_M}{L} wmc \tag{6}$$

Combining Eq. (2) with Eq. (4), the surface divergence of the $\xi$-vector could be derived as.

$$\begin{aligned} div_S \xi = &(1 - \frac{3}{5}\varepsilon_1 - \frac{17}{7}\varepsilon_2)(\frac{\partial n_x}{\partial x} + \frac{\partial n_y}{\partial y} + \frac{\partial n_z}{\partial z}) \\ &+ 12(\varepsilon_1 + 3\varepsilon_2)(n_x^2 \frac{\partial n_x}{\partial x} + n_y^2 \frac{\partial n_y}{\partial x} + n_z^2 \frac{\partial n_z}{\partial x}) \\ &- 3(\varepsilon_1 + 3\varepsilon_2)Q(\frac{\partial n_x}{\partial x} + \frac{\partial n_y}{\partial y} + \frac{\partial n_z}{\partial z}) \\ &- 3(\varepsilon_1 + 3\varepsilon_2)(n_x \frac{\partial Q}{\partial x} + n_y \frac{\partial Q}{\partial y} + n_z \frac{\partial Q}{\partial z}) \\ &+ 132\varepsilon_2(\frac{1}{n_x}\frac{\partial R}{\partial x} + \frac{1}{n_y}\frac{\partial R}{\partial y} + \frac{1}{n_z}\frac{\partial R}{\partial z}) \\ &- 132\varepsilon_2 R(\frac{1}{n_x^2}\frac{\partial n_x}{\partial x} + \frac{1}{n_y^2}\frac{\partial n_y}{\partial y} + \frac{1}{n_z^2}\frac{\partial n_z}{\partial z}) \\ &- 330\varepsilon_2(n_x \frac{\partial R}{\partial x} + n_y \frac{\partial R}{\partial y} + n_z \frac{\partial R}{\partial z}) \\ &- 330\varepsilon_2 R(\frac{\partial n_x}{\partial x} + \frac{\partial n_y}{\partial y} + \frac{\partial n_z}{\partial z}) \end{aligned} \tag{7}$$

where $n_x = \partial_x f_s / |\nabla f_s|$, $n_y = \partial_y f_s / |\nabla f_s|$ and $n_z = \partial_z f_s / |\nabla f_s|$, $Q = n_x^4 + n_n^4 + n_z^4$, $R = n_x^2 n_y^2 n_z^2$

To solve Eq. (7), the partial derivatives of solid fraction, such as: $\partial_x f_s$, $\partial_x^2 f_s$, $\partial_{x,y}^2 f_s$ …are needed to be accurately calculated. An actuate method to solve the partial derivatives of solid fraction for 2-D CA model is discussed in our previous work [20, 21]. In present article, the method is extended into 3-D application.

## 1.4 Cellular Automaton rule

For 3-D simulation, each interface cell has 6 neighbor cells in <100> directions, 12 neighbor cells in <110> directions and 8 neighbor cells in <111> directions. So, the neighborhood rule of the cell in 3-D is much more complicated than that in 2-D. It is very difficult to present a 3-D capture rule that has a lower mesh induced anisotropy. So in present work, the simplest capture rule - Von Neumann type (6 cells in <100> directions) is used, and a limited neighbor solid fraction (LNSF) method was used to control the mesh induced anisotropy.

The capture rule in 3-D CA model with LNSF method is described as follows: firstly, a limited neighbor solid fraction $fs_{LNSF}$ was set; secondly, the average solid fraction of the neighbor cells for each liquid cell $fs_{ave}$ was calculated. If the $fs_{ave}$ is larger than $fs_{LNSF}$ and this liquid cell should have a solidified neighbor cell in Von Neumann type, then it can be captured as an interface cell.

It is found that when $fs_{LNSF}$ gets smaller, the mesh induced anisotropy prefers <100> directions. Especially, if $fs_{LNSF}$ is equal to 0.0, the mesh induced anisotropy is the same as that caused by Von Neumann capture rule. However, when $fs_{LNSF}$ gets larger, the mesh induced anisotropy prefers <111> directions. The $fs_{LNSF}$ of 0.225 presents a balance between the two mesh induced anisotropies.

It can be seen that the capture rule in present 3-D CA model is a combination of the Von Neumann capture rule and the LNSF method. The Von Neumann capture rule makes sure that the CA model presents a sharp interface, and the LNSF method makes sure that all the liquid cells are captured under the similar conditions, thus the mesh induced anisotropy could be reduced.

## 1.5 Computational details

The equiaxed dendirtic growth from undercooled pure succinonitrile (SCN) melt is simulated using 3-D CA model, which is programmed using adaptive mesh refinement (AMR) technique. The thermal physical parameters of SCN used in the simulation are the same as reference [21].

The simulation domain is $0.128 \times 0.128 \times 0.128 mm^3$, and the mesh size is smaller than 0.5 μm. At the beginning, the central cell of computational domain is initialized as a solid cell and the states for the rest cells are set to liquid. The time step is calculated as follow.

$$\Delta t = 0.15 \frac{\Delta x^2}{\alpha} \tag{16}$$

where, $\Delta t$ is the time step, $\Delta x$ is mesh size.

The initial temperature and boundary temperature of the domain are set to $T_0$, thus the undercooling of the melt can be expressed as $\Delta T = T_m - T_0$.

## 2. Simulation results and discussion

### 2.1 The testing of mesh induced anisotropy

In order to test the mesh induced anisotropy, the morphologies under the isotropic interface free energy ($\varepsilon_1$=0.0, $\varepsilon_2$=0.0) were first simulated and the limited neighbor solid fraction $fs_{LNSF}$ was set to 0.225. Fig.1 shows the simulated morphologies grown from the pure SCN melt with the undercooling of 6K. Both of Fig.1 (a) and Fig.1 (b) used Von Neumann capture rule, and the differences between them are the use of LNSF method. It can be seen that when the LNSF method is used, the simulated morphology presents 24 primary branches with the same size evenly pointing at all the directions; and when the LNSF method is not used, all the 24 primary branches grew in <100> directions, which is caused by Von Neumann capture rule. It can be concluded that the LNSF method could reduce the mesh induced anisotropy.

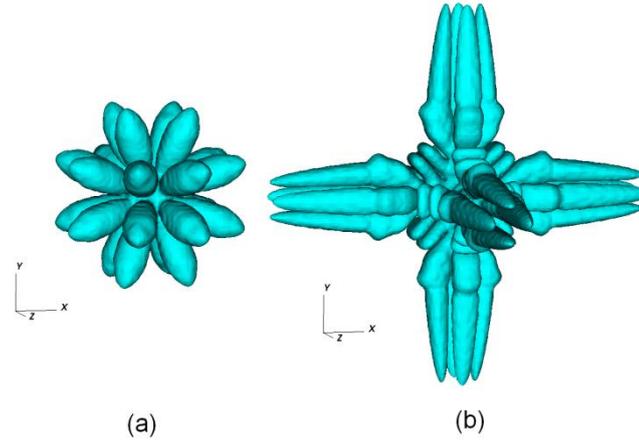

(a)　　　　　　　(b)

Fig. 1 Simulated solidifying morphologies in the undercooled pure SCN melt, $\Delta T$=6K, $\Delta x$ =0.5 μm, $\varepsilon_1$=0.0, $\varepsilon_2$=0.0: (a) with LNSF method; (b) without LNSF method

### 2.2 Dendrite morphology by 3-D CA model

Fig.2 shows the simulated morphologies grown from the pure SCN melt under the undercooling of 4.5K. The interface anisotropy parameters $\varepsilon_1$ and $\varepsilon_2$ are set to 0.10 and 0.0 for Fig.2 (a), 0.0 and -0.02 for Fig.2 (b) respectively. It can be seen that, a dendrite with the <100> preferred direction is grew with $\varepsilon_1$=0.10 and $\varepsilon_2$=0.0, and a dendrite with the <110> preferred direction is grew when the interface anisotropy parameters $\varepsilon_1$ and $\varepsilon_2$ are 0.00 and -0.02 respectively. Fig.2 (c) and Fig.2 (d) show the dendrite tips illustrated by AMR mesh. It can be seen that the solid-liquid interface is covered by the most refined meshes, which can describe the interface more accurately.

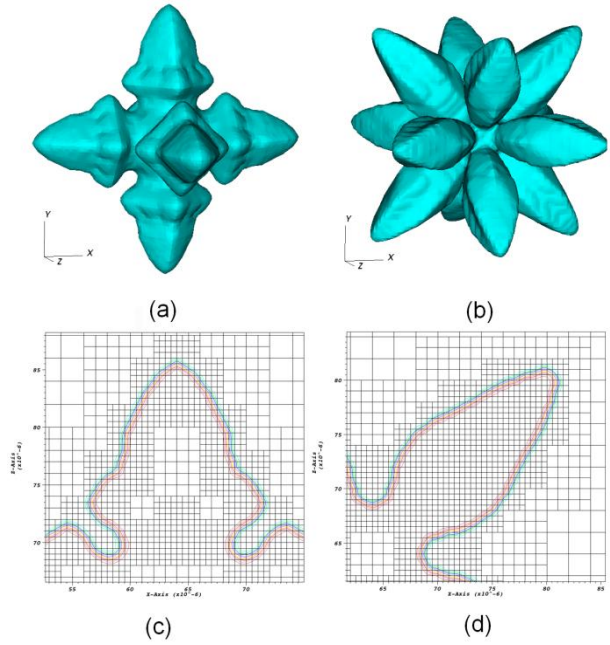

Fig. 2 Simulated morphologies during solidification from the undercooled pure SCN melt, $\Delta T=4.5K$, $\Delta x = 0.5$ μm: (a) $\varepsilon_1 = 0.10$, $\varepsilon_2=0.0$ (b) $\varepsilon_1 = 0.0$, $\varepsilon_2=-0.02$; (c) and (d) are the dendrite tips in 2-D AMR mesh for (a) and (b), respectively.

### 2.3 Dendrite orientation selection

The simulated morphologies by present 3-D CA model are further compared with the prediction of the spherical plot of the inverse of the interfacial stiffness, as expressed in Eq. (3), when the anisotropy parameter $\varepsilon_1$ changes from 0.0 to 0.16 with the fixed $\varepsilon_2$ of -0.02. It can be seen that the simulated morphologies by CA model are continuously changed with the $\varepsilon_1$ as expected from the minimum stiffness criterion.

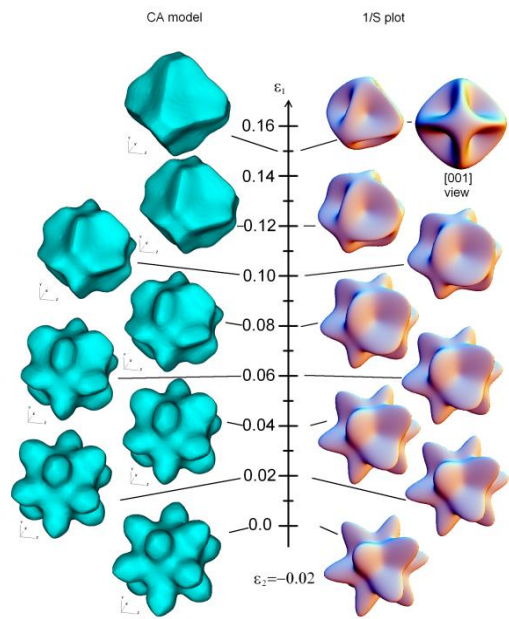

Fig. 3 The comparison of morphologies predicted by CA model and 1/S plot for different $\varepsilon_1$

with the $\varepsilon_2$ of -0.02 for the growth from undercooled pure SCN melt, $\Delta T=2.0K$, $\Delta x= 0.25$ μm

**Summary**

A 3-D cellular automaton model for describing the solidification of a pure substance was developed, in which a limited neighbor solid fraction (LNSF) method was presented to reduce the mesh induced anisotropy.

In order to simulate the orientation selection, an expansion description using two interface free energy anisotropy parameters ($\varepsilon_1$, $\varepsilon_2$) is used in present 3-D CA model. When $\varepsilon_1=0.10$, $\varepsilon_2=0.0$, the dendrite grew with the <100> preferred directions; and when $\varepsilon_1=0.0$, $\varepsilon_2=-0.02$, the dendrite grew with the <110> preferred directions.

The dendrite orientation selection between <100> and <110> dendrites were compared with the spherical plot of the inverse of the interfacial stiffness, using the different $\varepsilon_1$ with $\varepsilon_2=-0.02$. It is found that the simulated morphologies by present CA model are as expected from the minimum stiffness criterion.

**Acknowledgement**

This project was supported by the National Natural Science Foundation of China (Nos. 50971102 and 50901061) and the National Basic Research Program of China (No. 2011CB610402). The work was also supported by the Programme of Introducing Talents of Discipline to Universities (No.08040) and the fund of the State Key Laboratory of Solidification Processing in NWPU (Nos. 39-QZ-2009 and 02-TZ-2008).